\documentclass[prl,showpacs,twocolumn,preprintnumbers,amsmath,amssymb]{revtex4}

\usepackage{epsfig}
\usepackage{graphicx}
\usepackage{dcolumn}
\usepackage{color}
\usepackage{bm}

\begin{document}

\title{Interactions and superconductivity in heavily doped MoS$_2$}
\author{R. Rold\'an$^1$, E. Cappelluti$^{1,2}$, F. Guinea$^1$}

\affiliation{$^1$ Instituto de Ciencia de Materiales de Madrid. Consejo Superior de Investigaciones Cient{\'\i}ficas. Sor Juana In\'es de la Cruz 3, 28049 Madrid. Spain. \\
$^2$ Istituto de Sistemi Complessi, U.O.S. Sapienza, CNR, v. dei Taurini 19, 00185 Roma, Italy}

\begin{abstract}
We analyze the microscopic origin and the physical propertis
of the superconducting phase recently observed in MoS$_2$.
We show how the combination of the valley structure of the conduction band,
the density dependence of the  screening of the long range Coulomb interactions,
the short range electronic repulsion, and the relative
weakness of the electron-phonon interactions, makes possible the
existence of a phase where the superconducting order parameter has
opposite signs in different valleys,
resembling the superconductivity found in the pnictides and cuprates.
\end{abstract}
\maketitle

Molybdenum disulfide (MoS$_2$) is a layered semiconductor which
 can be exfoliated down to monolayer unit cells \cite{WS12}, like graphene \cite{Netal05,Netal05b,MCHH10}. The existence of an energy gap makes MoS$_2$ a convenient material for nanoelectronics \cite{Retal11,ZYMI12}. Metallic behavior can be induced, also like in graphene,
by means of electric field effects or by doping,
and the corresponding Fermi surface is typically made up
by inequivalent Fermi pockets \cite{K73,B78,Cetal87,Betal01,LE09},
defining a valley degree of freedom which is
strongly entangled with the spin degree of freedom \cite{Xetal12},
and it can be further controlled and manipulated, opening promising
perspectives for spintronics.
At high carrier concentrations ($n \sim 10^{14} {\rm cm}^{-2}$), and in the presence of high-$\kappa$ dielectrics, MoS$_2$ has also be shown
to undergo a superconducting transition,
with a doping-dependent critical temperature $T_c ( n )$ which exhibits
a maximum as function of $n$ and drops to zero at sufficiently large values of $n$ \cite{TMST12,Yetal12}.

A ferromagnetic behavior has been also reported in MoS$_2$
\cite{Zetal07,LZZC09,Metal12,Metal12b}, and it has been
related to edges or to the existence of defects \cite{VHN09,ASAC11}.
The magnetic properties of MoS$_2$ nanoribbons indicate that the electron-electron interactions are non negligible. The combination of significant electron-electron interactions and a two dimensional Fermi surface made up of many pockets is also
a hallmark of the cuprate and pnictide superconductors \cite{M10}, where the superconducting gap has a $d$-wave symmetry (cuprates)
or opposite sign in different pieces of the Fermi surfaces (pnictides) \cite{M10}.
A related gap structure has been also proposed for heavily doped graphene when the electron-electron interaction is sufficiently large \cite{GU12}.

In the present work we study the origin of superconductivity in heavily doped MoS$_2$, by considering the role of both electron-electron and electron-phonon interactions. We analyze first the general features of the effective interaction between charge carriers,
and we make semi-quantitative estimates of the strength of the different contributions to the effective coupling. We discuss next the competition between
the electron-electron and electron-phonon interactions, and the possible types of superconductivity that emerge.
Although a quantitative determination of the superconducting $T_c$
is outside the scope of our work,
the present analysis suggests that superconductivity in MoS$_2$ is
likely to be induced by the electron-electron interaction, and that a
superconducting phase with a non trivial gap structure
is possible.

\begin{figure}
\begin{center}
\includegraphics[width=1.\columnwidth]{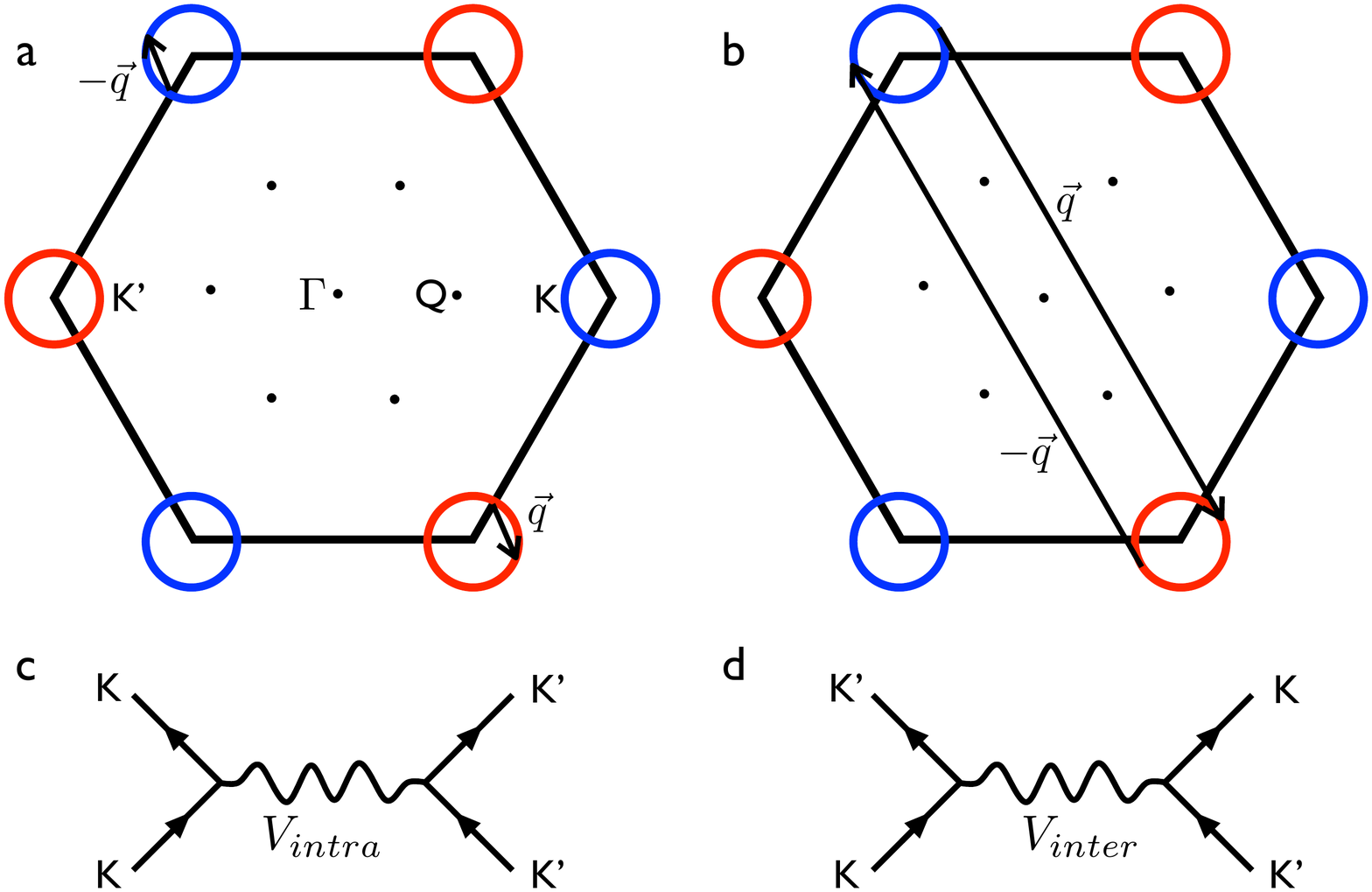}
\includegraphics[width=1.\columnwidth]{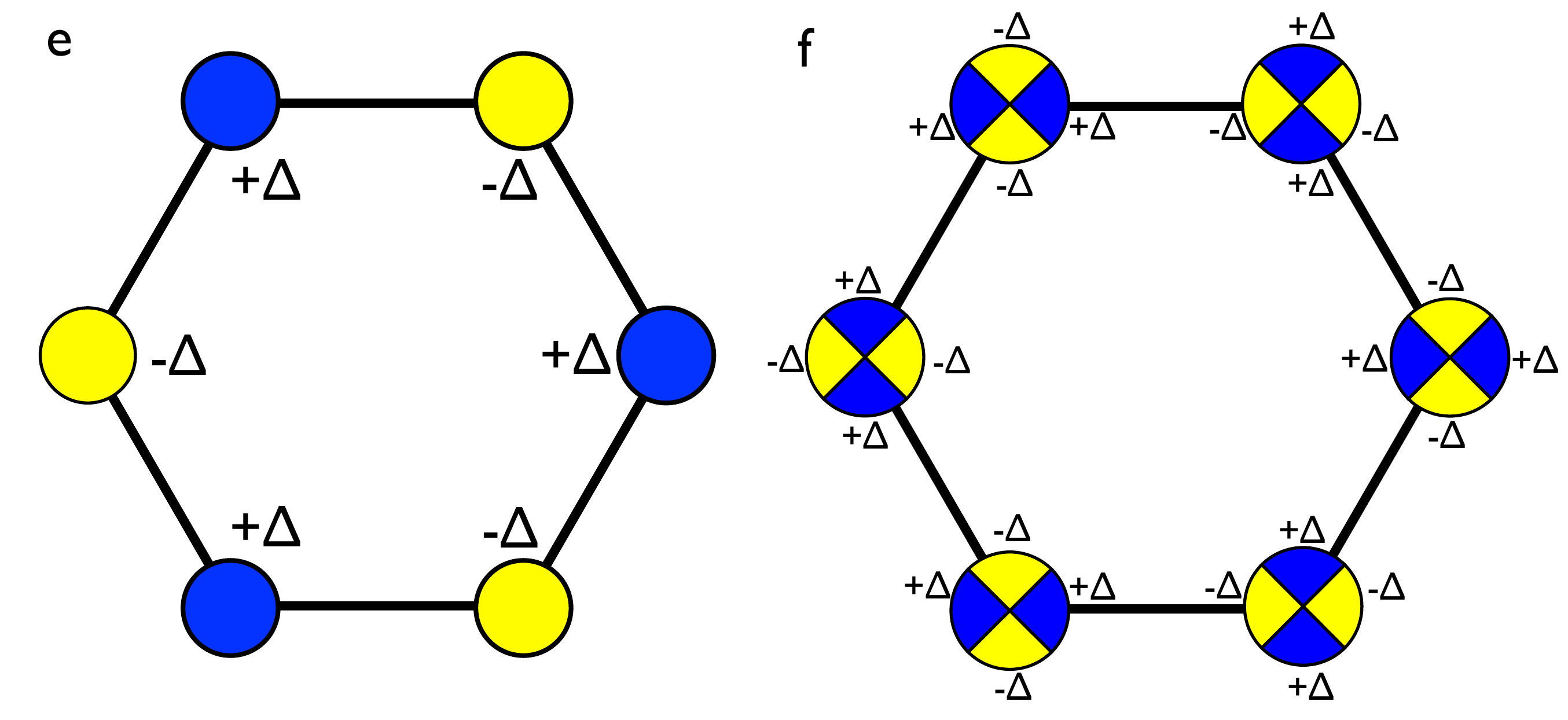}
\caption[fig]{\label{Fig:diagrams} Sketch of the intravalley (a)
and intervalley (b) scattering processes between Cooper pairs in
MoS$_2$.
Panels (c) and (d): corresponding Feynman diagrams associated to
intra- and
intervalley scattering. Full arrows are electron propagators, and wavy
lines are effective interactions.
The possible unconventional superconducting phases discussed in the text are sketched
in panels (e) and (f).}
\end{center}
\end{figure}

{\em Effective interactions.}
Following the experimental results of Ref. \cite{TMST12,Yetal12}, we assume that the carriers leading to the superconducting phase are electron like, confined in the first MoS$_2$ layer
closer to the high-dielectric gate.
The validity of this approximation
will be discussed later.
 We also assume that,
similar to monolayer MoS$_2$, the electron-like carriers
are located in the two inequivalent Fermi pockets
centered at the $K$ and $K'$ corners of the Brillouin Zone (see Fig. \ref{Fig:diagrams}), having thus a sizable $d$-orbital character with main $d_{3z^2-r^2}$ orbital
component.
At sufficiently large concentrations six additional inequivalent $Q$ valleys start to be filled, located half way between the  $\Gamma$ and $K$ point, with primary $d_{x^2-y^2}$ and $d_{xy}$ Mo orbital character. These secondary valleys at higher charge density
do not play a relevant role in our main discussion and will
be therefore neglected. Their possible effect will be however
addressed in the final discussion.

We explore here the possibility that
superconductivity is induced by effective electron-electron interactions,
associated with the direct Coulomb interaction between charge carriers or with the effective coupling induced by phonons.
As in Ref. \cite{GU12}, we consider singlet superconductivity.
Because of the time invariance symmetry, the Cooper pair electrons
(${\bf k}\uparrow,-{\bf k}\downarrow$)
reside in different valleys.
We can now classify,
as sketched in Fig. \ref{Fig:diagrams},
the interactions leading to scattering of the Cooper pairs
into intra- and intervalley couplings,
namely $V_{intra} ( \vec{\bf q} , \omega )$, $V_{inter} ( \vec{\bf q} , \omega )$, where $\vec{\bf q}$ and $\omega$ are the exchanged momentum and frequency.
We consider only scattering processes of carriers near the Fermi surfaces, which are assumed to be isotropic and centered at $K$ and $K'$, and we neglect the frequency dependence of the interaction.

The classification of the interaction in an intravalley and an intervalley
component allows us to define the dimensionless coupling constants \cite{GU12}
\begin{align}
\lambda_{\alpha} &= \rho ( \epsilon_F )
\int_0^\pi d\theta~ V_{\alpha}\left[2k_F \sin \left( \frac{\theta}{2} \right)\right] ,
\label{coupl}
\end{align}
where $\alpha$ labels the $intra$- and $inter$-valley scattering,
and $\rho ( \epsilon_F ) = m_{eff} / (2 \pi \hbar^2 )$ is the density of states
at the Fermi level per valley and per spin in terms of the effective mass $m_{eff}$.
Typical values are $m_{eff}\approx 0.5m_0$ \cite{KH12,PV12},  where $m_0$ is the free electron mass.
For realistic charge concentrations
the Fermi wavevector is much smaller than the dimensions of the Brillouin Zone,
$k_F \ll |  \vec{\bf K} |$, so that we can in good approximation
neglect the momentum dependence of $V_{intra} ( \vec{\bf q} ) , V_{inter} ( \vec{\bf q} )$, except for electron-electron intravalley scattering, as discussed below.

The existence of superconductivity requires $\lambda_{intra} \pm \lambda_{inter} < 0$, where the choice of the sign depends on whether the gaps in the two valleys have equal or opposite signs. The electron-phonon coupling leads to attractive interactions, $\lambda^{e-ph} < 0$, while the electron-electron couplings lead to repulsive interactions, $\lambda^{e-e} > 0$. However,
repulsive interactions can also lead to superconductivity provided
that $\lambda_{inter} > \lambda_{intra} > 0$.

{\em Electron-phonon interaction.} The electron-phonon interaction in single layer MoS$_2$ has been evaluated in Ref. \cite{KTJ12}, finding relevant contributions from three acoustic modes and from six optical modes.
Note that the coupling to the acoustic modes vanishes when the phonon wavevector approaches zero.  The leading couplings identified in Ref. \cite{KTJ12} are
thus the ones to the polar LO modes, $\hbar \omega_{LO} \approx 0.048$ eV, and to the homopolar mode, $\hbar \omega_{ho} \approx 0.05$ eV, where the S atoms oscillate out of the plane \cite{KTJ12}. The homopolar mode contributes to intravalley scattering, while the LO mode contributes to intravalley scattering, through the induced electric polarization, and also to intervalley scattering.

Using the notation in \cite{KTJ12}, we define an effective interaction,
at frequencies much smaller than the phonon energies, as
\begin{align}
V_{intra}^{LO} &= - \frac{g_{LO}^2}{\hbar \omega_{LO}} \times \Omega,, \nonumber \\
V_{inter}^{LO} &= - \frac{D_{LO}^2}{\hbar \omega_{LO}} \frac{\hbar}{2 M \omega_{LO}} \times \Omega, \nonumber \\
V_{inter}^{ho} &= - \frac{D_{ho}^2}{\hbar \omega_{ho}} \frac{\hbar}{2 M \omega_{ho}}  \times \Omega,
\label{v_e_eph}
\end{align}
where $g_{LO} \approx 0.098$ eV is the long wavelength polar coupling,
$D_{LO} \approx 2.6$ eV \AA$^{-1}$ and $D_{ho} \approx 4.1$ eV \AA$^{-1}$ are
deformation potentials for the LO and homopolar modes,
respectively, and $\Omega$ is the area of the unit cell.
We take for $M$ the mass of the sulfur atom,
which is much lighter than Mo, and thus expected to be dominant.
Adding the three contributions, we find thus:
\begin{align}
\lambda_{intra}^{ph} &\approx - 0.36, \nonumber \\
\lambda_{inter}^{ph} &\approx - 0.13,
\label{l_e_ph}
\end{align}
which account for the respective intravalley and intervalley
electron-phonon coupling constants.

{\em Electron-electron interaction.}
After having estimated the electron-phonon coupling,
we address now the electron-electron repulsive interaction.
Intravalley scattering is operative only for small momenta/large distance,
where, as in Ref. \cite{GU12},
the electron-electron interaction is determined
by the screened Coulomb potential:
\begin{align}
V_{intra}^{e-e} ( q ) &= \frac{2 \pi e^2}{\epsilon_0 ( q + q_{FT} )},
\label{v_intra_e_e}
\end{align}
where $\epsilon_0$ is the dielectric constant of the environment, and $q_{FT} = 2 \pi e^2 \rho ( \epsilon_F ) / \epsilon_0$ is the Thomas-Fermi wave-vector.

\begin{figure}
\begin{center}
\includegraphics[width=0.9\columnwidth]{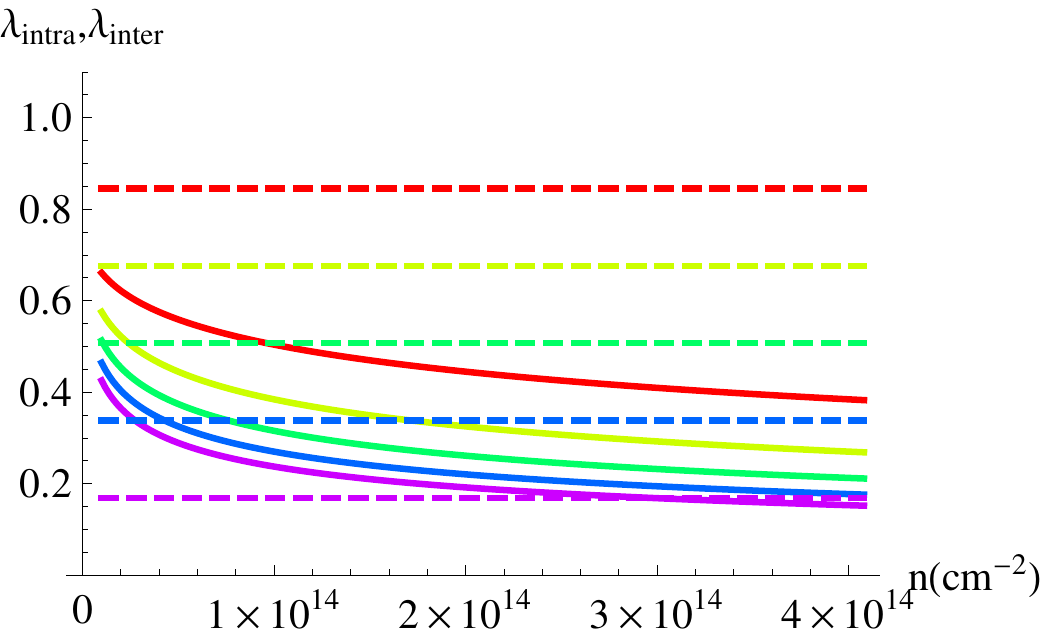}
\caption[fig]{\label{lambda_e_e} Intravalley and intervalley
dimensionless couplings $\lambda_{intra}^{e-e}$,
$\lambda_{inter}^{e-e}$
due to electron-electron interactions as
functions of density $n$, for different values of $\epsilon_0$ and $U_{4d}$.
Full lines: $\lambda_{intra}$. (from top to bottom:
$\epsilon_0 = 10 , 20 , 30, 40 , 50$);
broken lines: $\lambda_{inter}$ (from top to bottom: $U_{4d} = 10$ eV, 8 eV, 6 eV,
  4 eV, 2 eV).}
\end{center}
\end{figure}

On the other hand,
the contribution of the electron-electron interaction to intervalley scattering
is associated with the short range part of the Coulomb potential. The leading term in this interaction is thus given by the  Hubbard term, namely
the repulsion between two electrons with opposite spin in the same atomic orbital.
Since, as we discussed above, electronic states close to the $K$ and $K'$ points
have a dominant Mo $4d$ character, we can therefore approximate:
\begin{align}
V_{inter}^{e-e} &\approx U_{4d } \times \Omega,
\label{v_inter_e_e}
\end{align}
resulting thus in a ${\bf k}$-independent and density-independent interaction.

\begin{figure}
\begin{center}
\includegraphics[width=0.9\columnwidth]{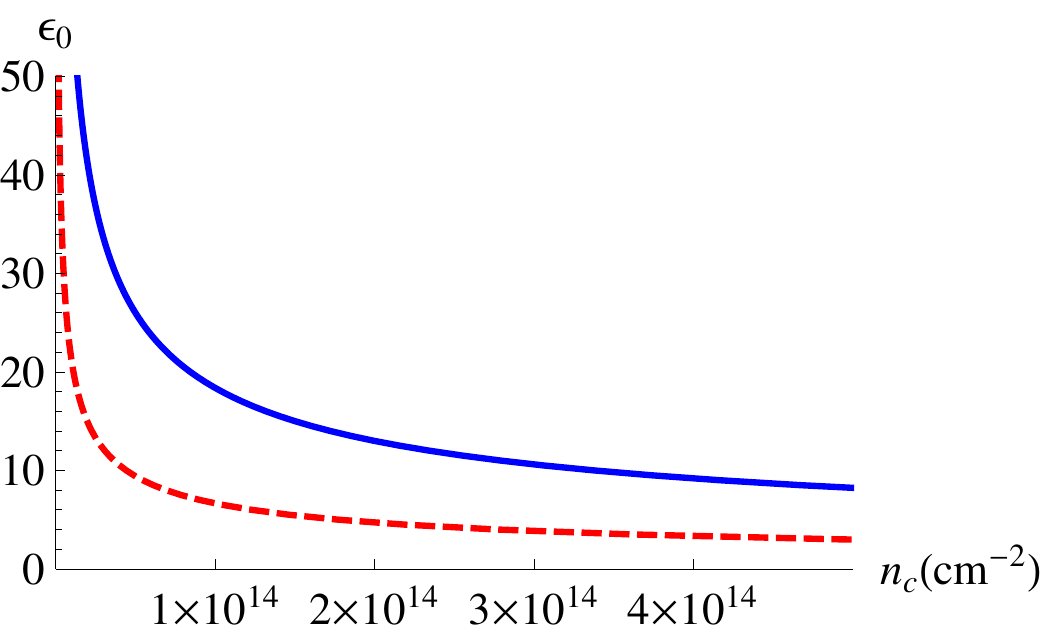}
\caption[fig]{\label{n_c} Relation between the critical carrier density, $n_c$,
required for the existence of superconductivity and the dielectric constant of the environment, $\epsilon_0$.
Solid blue curve: $U_{4d} = 2$ eV, dashed red curve: $U_{4d} = 4$ eV.}
\end{center}
\end{figure}

{\em Results.}
We discuss now the consequences of the electron-phonon vs. electron-electron
and the intra- vs. intervalley interaction in regards to the superconducting order.
Along this line, unfortunately,
the values of the dielectric constant, $\epsilon_0$, and of
the Hubbard term, $U_{4d}$, are not well known. The experiments reported in \cite{TMST12,Yetal12} are done in the presence of a dielectric with a high value of $\epsilon_0$.
The ionization energies \cite{MoS2properties} of the Mo atom
allow us to make an order of magnitude estimate of $U_{4d}$. Alternatively, we can calculate the Coulomb integral for the $4d$ orbitals of Mo, or replace the orbital by a charged sphere with the same radius.
All these different approaches converge on the order of magnitude
of the Hubbard repulsion, but a more accurate determination
of its specific value is lacking.
In this situation, we will consider $U_{4d}$ and $\epsilon_0$ as a free variable
parameters and we will investigate their effects on
the superconducting properties of MoS$_2$ in the range
$\epsilon_0 \approx 10 \div 50$ and $U_{4d} \approx 2 \div 10$ eV.

Fig.~\ref{lambda_e_e} shows the dependence
of the intravalley and intervalley electron-electron coupling constants
$\lambda_{intra}^{e-e}$ and $\lambda_{inter}^{e-e}$ on
the specific values of $\epsilon_0$ and $U_{4d}$.
In particular, the intravalley coupling $\lambda_{intra}^{e-e}$ shows an initial
significant dependence on density, due to the screening increase with density
for $q_{FT} \lesssim k_F$, wheeras $\lambda_{intra}^{e-e}$ saturates for $q_{TF} \gg k_F$.
A comparison with the corresponding coupling constants for electron-phonon interaction, provided by Eq. (\ref{l_e_ph}), shows that the values of $\lambda_{intra}^{e-e}$ and $\lambda_{inter}^{e-e}$ in the limit of high dielectric constant and moderate $U_{4d}$ are larger in this regime than those of $\lambda_{intra}^{e-ph}$ and $\lambda_{intra}^{e-ph}$, suggesting thus that superconductivity is due to the electron-electron interaction.
Furthermore, since  $\lambda_{inter} = \lambda_{inter}^{e-ph} +
\lambda_{inter}^{e-e} > 0$,
the superconducting phase is expected to have gaps
with opposite signs in the two valleys, as sketched in Fig. \ref{Fig:diagrams}e.
Note that superconductivity of this type is possible only
when the carrier density $n$ satisfies
the condition $\lambda_{intra} ( n ) - \lambda_{inter} < 0$. This inequality defines a critical density $n_c$ above which superconductivity appears. The dependence of $n_c$ on the dielectric constant of the environment is shown in Fig.~\ref{n_c}.

Apart from the more likely symmetry of the gap discussed above, it is
worth to notice that an additional superconducting phase is possible
due to the modulation of the electron-electron interaction due to
screening. This mechanism always leads to an order parameter with a
modulated $k$-dependent gap within each valley \cite{KL65}
(see Fig. \ref{Fig:diagrams}f).
However, the resulting critical temperatures for this phase are typically
very low \cite{GU12}, so that it is very unlike to be related to
the experimental evidence of superconductivity in this material.

{\em Superconducting phase.} The change in sign of the superconducting gap in the two valleys implies that the superconducting phase in MoS$_2$ have unusual properties
with respect to conventional superconductors.
More specifically we can remind \cite{KL65}:
i) Elastic scattering is pair breaking, leading to the suppression of superconductivity when $v_F / \ell \gtrsim | \Delta |$, where $v_F =  \hbar k_F  / m_{eff}$ is the Fermi velocity, $\ell$ is the elastic mean free path, and $\Delta$ is the superconducting gap; ii) Strong scatterers induce localized Andreev states within the gap; iii) Andreev states also appear at certain edges. However, unlike graphene,
it should be reminded taht MoS$_2$ also shows a significant spin-orbit coupling,
so that the combination of a non trivial superconducting order parameter with
the spin-orbit coupling can lead to other interesting properties.

{\em Open questions.}
An intriguing role, in the above scenario,
is played by the specific value of the effective mass $m_{eff}$.
Apart from the determination of the effective couplings,
such parameter is relevant in assessing the robustness
of the Fermi surface structure depicted in Fig. \ref{Fig:diagrams}.
The previous analysis was  based on the
assumption of an electronic structure similar to thant in monolayer MoS$_2$,
where the carriers occupy two inequivalent valleys
corresponding at the absolute minima in the conduction band
at the $K$ and $K'$ points. However, it should be reminded that
six inequivalent secondary minima are also predicted at the Q points,
midway between the $\Gamma$ and the $K$ and $K'$ points, shown by the dots in Fig. \ref{Fig:diagrams}.
In bilayer and multilayer MoS$_2$ the minima at the $Q$ points
are expected to lie below the valleys at the $K$ and $K'$ points, and in monolayer MoS$_2$
the minima in $Q$ start to be filled at sufficiently high carrier densities, once the pockets in $K$ and $K'$ are partially filled.
Since the relative filling of the pockets at $K$-$K'$ and $Q$ points depends
on the corresponding effective masses, an accurate determination
of $m_{eff}$ could also assess the possible presence
of Fermi pockets at the $Q$ points.
It should be also noted that the value
of the effective mass $m_{eff}$ determines the screening properties and, as a result,
the distribution
of the gate-induced charge in the multilayer MoS$_2$.
In Ref. \cite{Yetal12} it was suggested that,
because of the strong electric field,
carriers are mainly localized in the first layer close to the high-dielectric gate
 (note that the high carrier densities enhance the screening and
thus the confinement).

\begin{figure}
\begin{center}
\includegraphics[width=0.9\columnwidth]{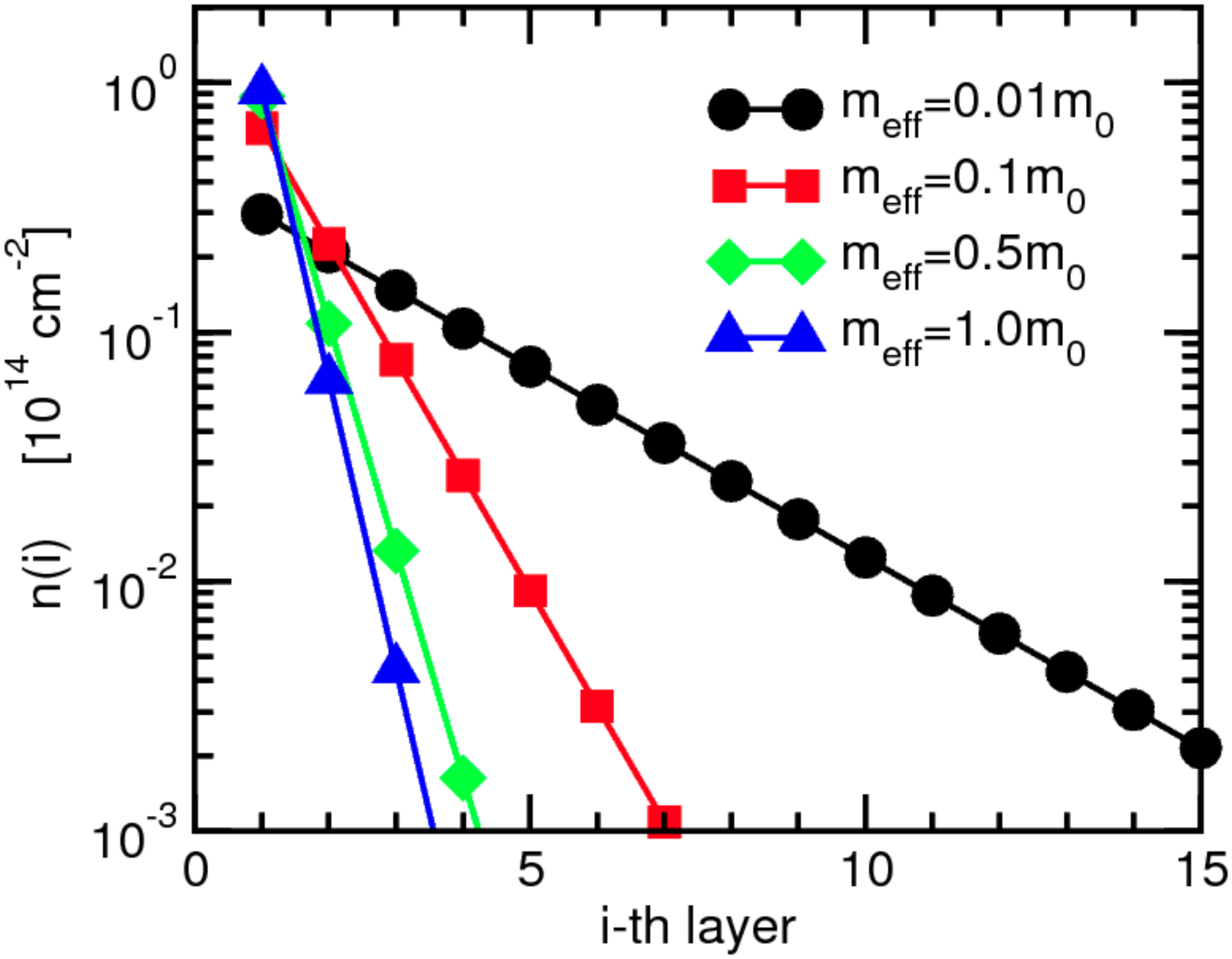}
\caption[fig]{\label{screening} Distribution of the electric charge
  per layer $n(i)$ in multilayer MoS$_2$ under gated conditions
for $n=10^{14}$ cm$^{-2}$, for different values of the effective mass.}
\end{center}
\end{figure}

Theoretical first-principles-based calculations indicate
a mass $m_{eff}$ of the order $m_{eff} \approx 0.4-1 m_0$
\cite{K73,H74,B78,Cetal87,Betal01,LE09,PV12,KH12},
whereas early experimental measurements suggest
$m_{eff}$ ranging from $ \sim 0.01m_0$ \cite{EY67} to  $\sim 1 m_0$ \cite{K76,Metal77,FR75}.
In the absence of a definitive estimate of $m_{eff}$,
we checked the dependence of the induced charge distribution
on $m_{eff}$ by performing a Thomas-Fermi-like
calculation using a similar model for the screening of a MoS$_2$ multilayer
as in Ref. \cite{Cetal12}.
The results for $n \approx 10^{14} {\rm cm}^{-2}$ 
using different possible values of $m_{eff}$,
are shown in Fig.~\ref{screening}. They confirm that for
$m_{eff} \gtrsim 0.1 m_0$ the charge is indeed  confined
only in the layer closest to the high-$\kappa$ gate \cite{Yetal12}.
Note that, since the charge is strongly concentrated
in the first layer, an appropriate generalization
of the model on a discrete layer distribution was here employed,
whereas the continuum original model of Ref. \cite{Cetal12}
would be inappropriate.
Our results corroborate thus the
initial assumption that the single layer model is valid and only
the $K$ and $K'$ valleys are occupied for the densities that lead to superconductivity in \cite{TMST12,Yetal12}.

It is worth to mention however that,
exotic superconducting phases,
with a gap with different signs in different valleys as sketched in Fig. \ref{Fig:diagrams}e,
are expected to appear
also in the cases when only the Fermi pockets at the $Q$ points are filled
(expected in the low density regime when the carriers are spread among a few layers)
or even in the case at higher fillings where both pockets ($Q$ and $K$) are occupied.
Such exotic phases would share a similar phenomenology
as in the case here considered of only two valleys at the $K$ and $K'$ points,
in particular in regards to the features (i)-(iii) mentioned above.

As final consideration, it should be remarked that our analysis implies that 
that the main dependence of the superconducting phase
on the charge density is by the suppression of the long range Coulomb repulsion,
due to the higher screening at higher charge concentrations (see Fig.~\ref{lambda_e_e}).
Hence, our analysis at this level does not account for the decrease of $T_c$
observed at high carrier densities $n \gtrsim 1.3 \times 10^{14} {\rm cm}^{-2}$ \cite{Yetal12}. A possible explanation for this behavior
could be the change in the orbital character of the states close to $K$ and $K'$ point,
which loose $4d$ orbital character at higher doping, resulting
in a weaker Hubbard-like intervalley interaction, or the change in the topology of the Fermi surface, as the $Q$ valleys start to be filled.

{\em Conclusions.}
In this paper we
have analyzed the appearing of a superconducting phase of MoS$_2$ at high carrier concentrations and for strong screening of the long range Coulomb potential. The significant short range repulsion between carriers at the conduction band allows for a superconducting phase induced by the electron-electron interaction, where the gap
acquires opposite signs in the two inequivalent pockets of the conduction band. This superconducting state, similar to that found in the cuprate and pnictide superconductors, is expected to show interesting topological features, such as Andreev states at edges and grain boundaries.

{\em Acknowledgements.} F. G. acknowledges financial support from MINECO, Spain, through grant FIS2011-23713, and the European Union, through grant 290846. R. R. acknowledges financial support
from the Juan de la Cierva Program (MINECO, Spain). E.C. acknowledges the Marie Curie grant PIEF-GA-2009-251904.
\bibliography{MoS2_superconductivity}

\end{document}